\documentclass[a4paper,12pt]{revtex4} 
  
\usepackage[margin=2.75cm]{geometry}

\usepackage{graphicx}
\usepackage{epsfig}
\usepackage{epstopdf}

\usepackage{hyperref}

\newcommand{\wprime}{\ensuremath{W^\prime}~}
\begin{document}

\def \a'{\alpha'}
\baselineskip 0.65 cm
\begin{flushright}
\ \today
\end{flushright}

{\vskip 0.5 cm}
\begin{center}{\large
{\bf The s-Channel Single Top Quark Production as a Constraint for $W'$ Boson Contribution}} {\vskip 0.5 cm} {\bf ${\rm Seyed~ Yaser~ Ayazi}$$^{1,\dagger}$ and ${\rm Saeid~ Paktinat~ Mehdiabadi}$$^{2,3,\ddagger}$}{\vskip 0.5 cm
}
{\small $^1$$Physics~Department$,~ $Semnan~University$,~$P.O.~Box.~35131-19111,~Semnan$,~$Iran$
$^2$$School~of~Particles~and~Accelerators,~Institute~for~ Research~in~Fundamental$
$Sciences~(IPM),~P.O.~Box~19395-5531,~Tehran,~Iran$
$^3$$Faculty~of~Physics,~Yazd~University,~P.O.~Box~89195-741,~Yazd,~Iran$
$^\dagger$syaser.ayazi@semnan.ac.ir; $^{\ddagger}$paktinat@ipm.ir}

\end{center}
\thispagestyle{empty}

\begin{center}
	{\bf Abstract}
{\vskip 0.5 cm}
\begin{minipage}{13cm}
An analysis is performed to constrain the $W'$ boson production using the measurement of the s-channel single top quark production cross section. Both phenomenological and statistical approaches are examined and results are presented. In the best case,  $W'$ bosons that interact only to the right-handed fermions are excluded below 1390 GeV. To our knowledge, it is the first time that the measured cross section of the s-channel single top quark production from the colliders is used to rule out part of the phase space of the $W'$ boson.
\end{minipage}
\end{center}

{\vskip 0.5 cm}

\section{Introduction}
\label{sec:intro}
The existence of a new massive charged gauge boson, known as \wprime boson, is proposed by many new physics scenarios. 
The minimal extension of the Standard Model (SM) gauge group incorporating \wprime is known as $SU(2) \times SU(2) \times U(1)$ (G(221)) models \cite{Hsieh:2010zr}.
The Kaluza-Klein excitations of the SM $W$ boson in extra dimension models is another famous example \cite{Burdman:2006gy}. 
In this paper, a general description of this new massive charged particle interactions is considered, without constraining the couplings to a special model. 

The general Lagrangian describing the fermionic interactions of \wprime boson can be written as:
\begin{equation}
{\cal L} = \frac{V^{\prime}_{ij}g_{W}}{2\sqrt{2}} \overline{f}_i\gamma_\mu \bigl(
a_R (1 + \gamma^5) + a_L (1 - \gamma^5) \bigr) W^{\prime\mu} f_j +
\mathrm{H.c.} \,, \label{eq:genlan}
\end{equation}
As for the notations, we follow the definitions of the Ref.~\cite{Sullivan:2002jt} and \cite{Duffty:2012rf} with small modifications, where $V^{\prime}_{ij}$ is a $3\times3$ identity matrix for leptons or the CKM matrix for quarks, $g_W = e/sin(\theta_W)$ is the SM weak coupling constant and $a_{R,L}$ are the strengths of the right and left couplings. Through this work, $a_{R,L}$ are assumed to be real values.

There have been many direct searches for \wprime in the high energy particle colliders, but up to now, all of them have failed. The searches include the 
fully left-handed \wprime ($a_{R}$ = 0) when $a_{L}$ is not constrained or when  $a_{L}$ = 1, known as the Sequential SM (SSM) \cite{Altarelli:1989ff}. The CMS experiment at the CERN LHC \cite{Chatrchyan:2008aa} has excluded the SSM \wprime with masses below 4.1 TeV \cite{Khachatryan:2016jww} at 95\% confidence level (CL) by looking at the tail of the transverse mass distribution of a lepton which comes from the decay of a \wprime associated with missing transverse energy coming from a neutrino. The search uses 2.3 fb$^{-1}$ of proton-proton (pp) collisions in the center of mass energy of 13 TeV. The ATLAS experiment \cite{Aad:2008zzm} in a similar search  uses 36.1 fb$^{-1}$ of data and rules out the SSM \wprime below 5.1 TeV \cite{Aaboud:2017efa}. Another class of the searches, consider the 
fully right-handed \wprime ($a_{L}$ = 0), where decay to leptons are
either closed or highly suppressed due to introduction of heavy right-handed neutrinos \cite{Khachatryan:2016jqo,Aad:2014xea}. Search for decay of \wprime	to light jets suffers from the high level of the QCD multijets backgrounds, but if \wprime is heavier than ~180 GeV, it can decay to a pair of top and bottom quarks (tb = t$\bar{\rm b}$ or $\bar{\rm t}$b) which has a distinguished signature due to the possibility of tagging the jets originating from the b quarks. Both CMS and ATLAS experiments have looked at this final state in different center-of-mass energies. The most recent result from the ATLAS experiment  is for $\sqrt{s}$ = 8 TeV that excludes the \wprime ~with masses below 1.92 TeV in 95\% CL \cite{Aad:2014xea}. The CMS experiment rules out the right-handed \wprime ~lighter than 2.6 TeV \cite{Sirunyan:2017ukk}, by using up to 2.6 fb$^{-1}$ of 13 TeV data, in the same channel. The previous limit from the CMS experiment, in this channel, was 2.15 TeV based on 19.7 fb$^{-1}$ of 8 TeV data \cite{Khachatryan:2015edz}.

When \wprime decays to tb, the final state is very similar to the final state of the s-channel single top  production, where $W$ boson is the off-shell mediator to produce top and bottom quarks. If \wprime exists, it can affect the cross section of the SM s-channel single top production, so measuring this cross section can constrain the \wprime contribution. The contribution of  \wprime boson to s-channel single top production is discussed in some other papers also \cite{Boos:2006xe,Drueke:2014pla,Tait:2000sh,Ayazi:2010jd}, but it is the first time that the measured cross section of s-channel single top is used to constrain the \wprime contribution. 

The s-channel production of the single top quark was first observed at the Tevatron experiments in $\sqrt{s}$ = 1.96 TeV with the statistical significance of 6.3 standard deviations ($\sigma$) \cite{CDF:2014uma}, but the most accurate measurement of the cross section of this process is done by the ATLAS experiment \cite{Aad:2015upn}. The measured cross section is $4.8  \pm 0.8~(\rm stat.)^{+1.6}_{-1.3}~(\rm syst.)$ with a signal significance of 3.2 $\sigma$ in 20.3 fb$^{-1}$ of the data in $\sqrt{s}$ = 8 TeV. 

It is shown \cite{Boos:2006xe} that only the left-handed \wprime can interfere with the SM $W$ boson contribution and its effect on the total cross section is not too large, although the shape of the partial cross sections can be changed dramatically. The right-handed \wprime does not have any interference, because the interactions of the SM $W$ boson are only left ($V-A$) type. 

In the next section, the phenomenology of the \wprime boson is reviewed shortly and the 
measured cross section of the s-channel production of the single top quark is used to constrain the \wprime contribution. In continue, the raw data reported by the LHC experiments are used to constrain the \wprime contribution by a statistical method.

\section{\wprime boson Phenomenology}
In this section, we study the effects of \wprime  boson
on the cross section of single top production and discuss the  possibility of detecting \wprime boson at LHC. As it is mentioned in the previous section,  in order to calculate the cross section of $pp\rightarrow
tb$, we consider the most general effective \wprime Lagrangian describing the interaction with the SM fermions. The cross section of $pp\rightarrow tb$ is given by:
\begin{eqnarray}
\sigma(pp\rightarrow tb) & = &\sum_{qq'} \int
dx_1dx_2q(x_1){\overline q'}(x_2)
\widehat{\sigma}(q{\overline q'}\rightarrow tb), \
\end{eqnarray}
where $q(x_i)$ and $\overline{q}'(x_i)$ are the parton distribution functions of  quarks. $x_1$ and $x_2$ are the parton momentum
fractions. The partonic cross section takes the form
\cite{Boos:2006xe}:
\begin{eqnarray}\label{eq:cross}
\label{formula_sigma} \widehat{\sigma}=\sum_{q,q'} \frac{\pi
	\alpha_W^2}{6}|V_{tb}|^2 |V_{qq'}|^2 \frac{(\hat{s}-M_t^2)^2 (2
	\hat{s}+M_t^2)}{\hat{s}^2} \left[
\frac{1}{(\hat{s}-m_W^2)^2+\Gamma_W^2 m_W^2} + \right.
\\ \nonumber +
\frac{2a^2_L((\hat{s}-m_W^2)(\hat{s}-M_{W'}^2)
	+\Gamma_W^2\Gamma_{W'}^2)} {((\hat{s}-m_W^2)^2+\Gamma_W^2
	m_W^2)((\hat{s}-M_{W'}^2)^2 +\Gamma_{W'}^2 M_{W'}^2)}
+  \\ \nonumber \left. + \frac{a_L^4+a_R^4+2a_L^2a_R^2}
{(\hat{s}-M_{W'}^2)^2+\Gamma_{W'}^2 M_{W'}^2} \right]
\end{eqnarray}
where $\Gamma_W$ is the width of $W$ boson, $\alpha_W =
g_W^2/(4\pi)$ and $\hat{s}=x_1 x_2S$ is the parton center of mass
energy while $S$ is the $pp$ center of mass energy. Width of the
\wprime is given in \cite{Boos:2006xe}.

\begin{figure}[!htb]
	\begin{center}
		\centerline{\hspace{0cm}\epsfig{figure=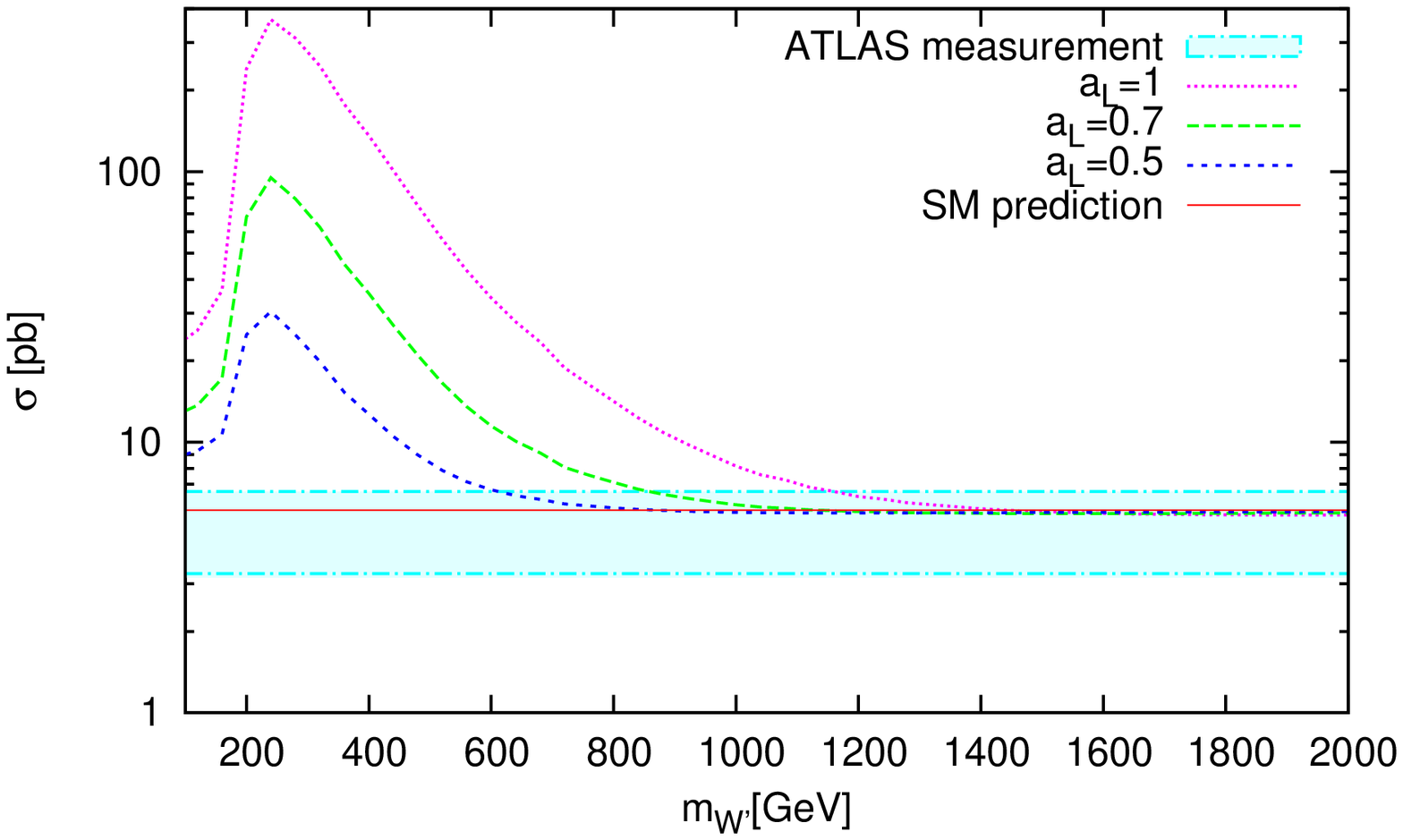,width=6.5cm}\hspace{0cm}\epsfig{figure=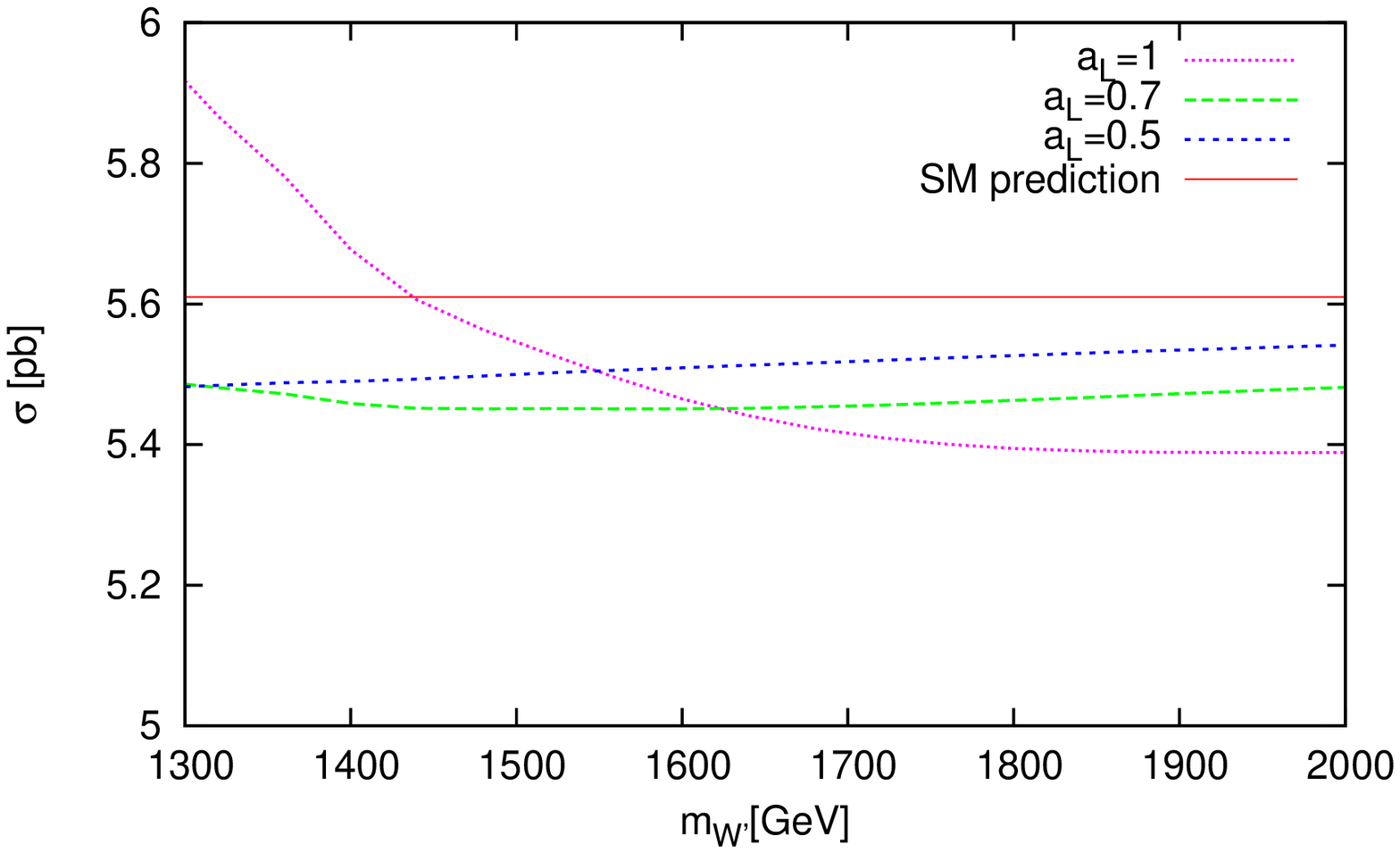,width=6.5cm}}
		\centerline{\vspace{-1.5cm}\hspace{0.5cm}(a)\hspace{6cm}(b)}
		\centerline{\vspace{-0.0cm}}
	\end{center}
	\caption{a) The total cross section of tb production as a function of the \wprime mass for several values of $a_L$ while $a_R=0$. b) To observe distractive effect of $a_L$, we zoom in the region  $m_{W'}=1300$ to $2000 \rm~ GeV$.}\label{fig:totalcross}
\end{figure}

As it is mentioned, the most precise measurement of single top production has been achieved by the ATLAS collaboration \cite{Aad:2015upn}.    
In Fig.~\ref{fig:totalcross}-a, we display the total cross section of
$pp\rightarrow tb$ versus $M_{W'}$ for several values of $a_L$ while $a_R=0$. To calculate
$\sigma(pp\rightarrow tb)$, we have used the CTEQ6.6M
parton distribution function \cite{Lai:2010nw}.  We have set $\sqrt{S}=8~ \rm TeV$  and a
SM cross section of $5.61~\rm pb$ for the
$\sigma(pp\rightarrow tb)$ was considered \cite{Aad:2015upn}. The 
horizontal line  depicts SM prediction for this
process. The shadow cyan area shows the allowed bound which is consistent with the ATLAS measurement.  As it is seen in Eq.~\ref{eq:cross}, for $a_R=0$ and non-zero $a_L$,  effect of \wprime exchange can be destructive.
To better show this effect, we zoom in the high \wprime mass region, where this effect can be important. In Fig.~\ref{fig:totalcross}-a, 
there exists a peak at areas which square of center of mass energy is close to the squared total mass of top and bottom quarks.   

\begin{figure}[!htb]
	\begin{center}
		\centerline{\hspace{0cm}\epsfig{figure=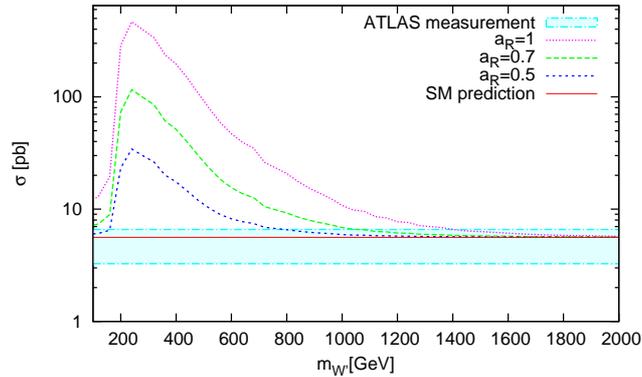,width=8.5cm}}
		\centerline{\vspace{-2.0cm}}
	\end{center}
	\caption{The total cross section of tb production as a function of the \wprime mass for several values of $a_R$ while $a_L=0$.}\label{fig:totalcross1}
\end{figure}

\begin{figure}[!htb]
	\begin{center}
		\centerline{\hspace{0cm}\epsfig{figure=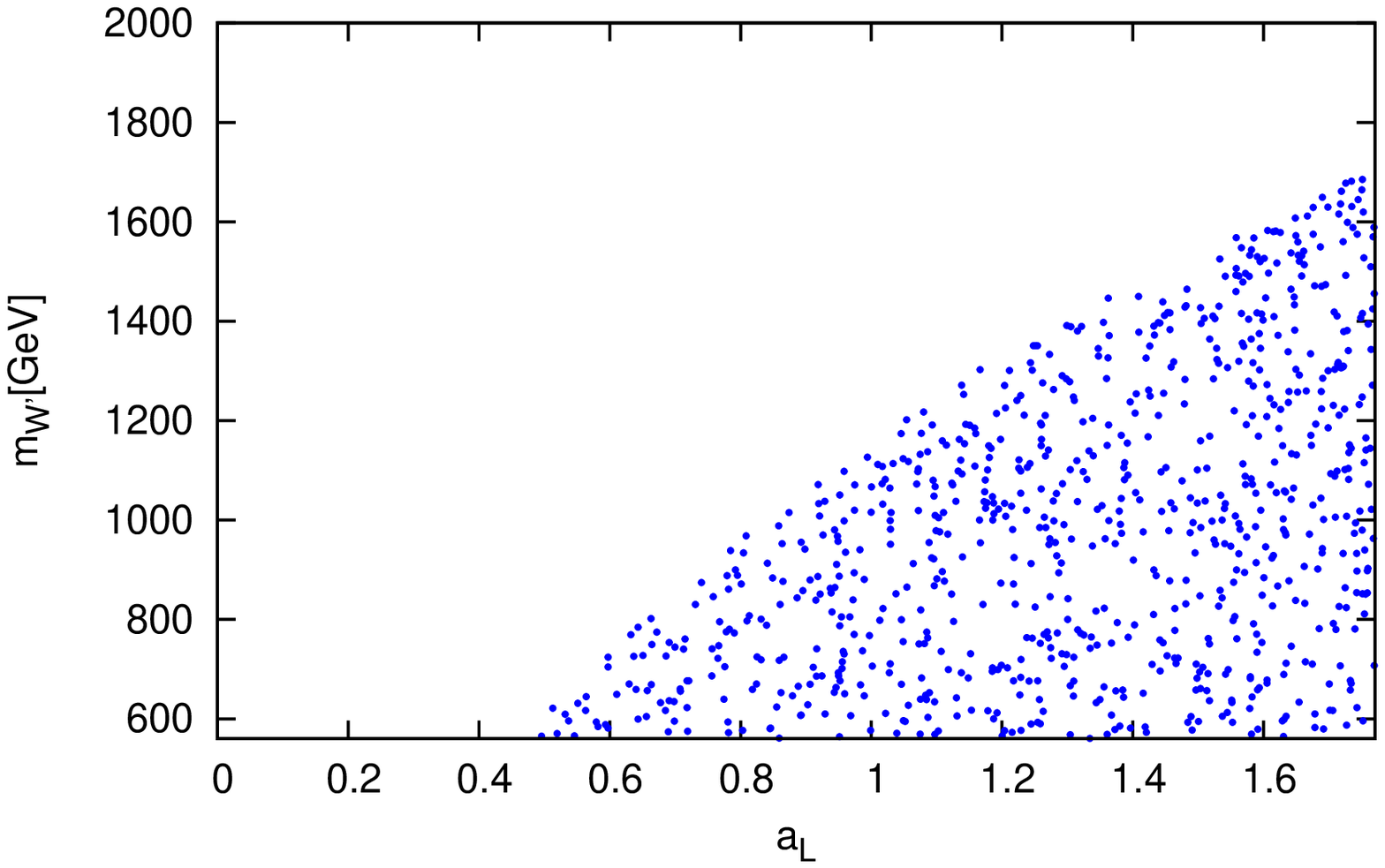,width=6.5cm}\hspace{0cm}\epsfig{figure=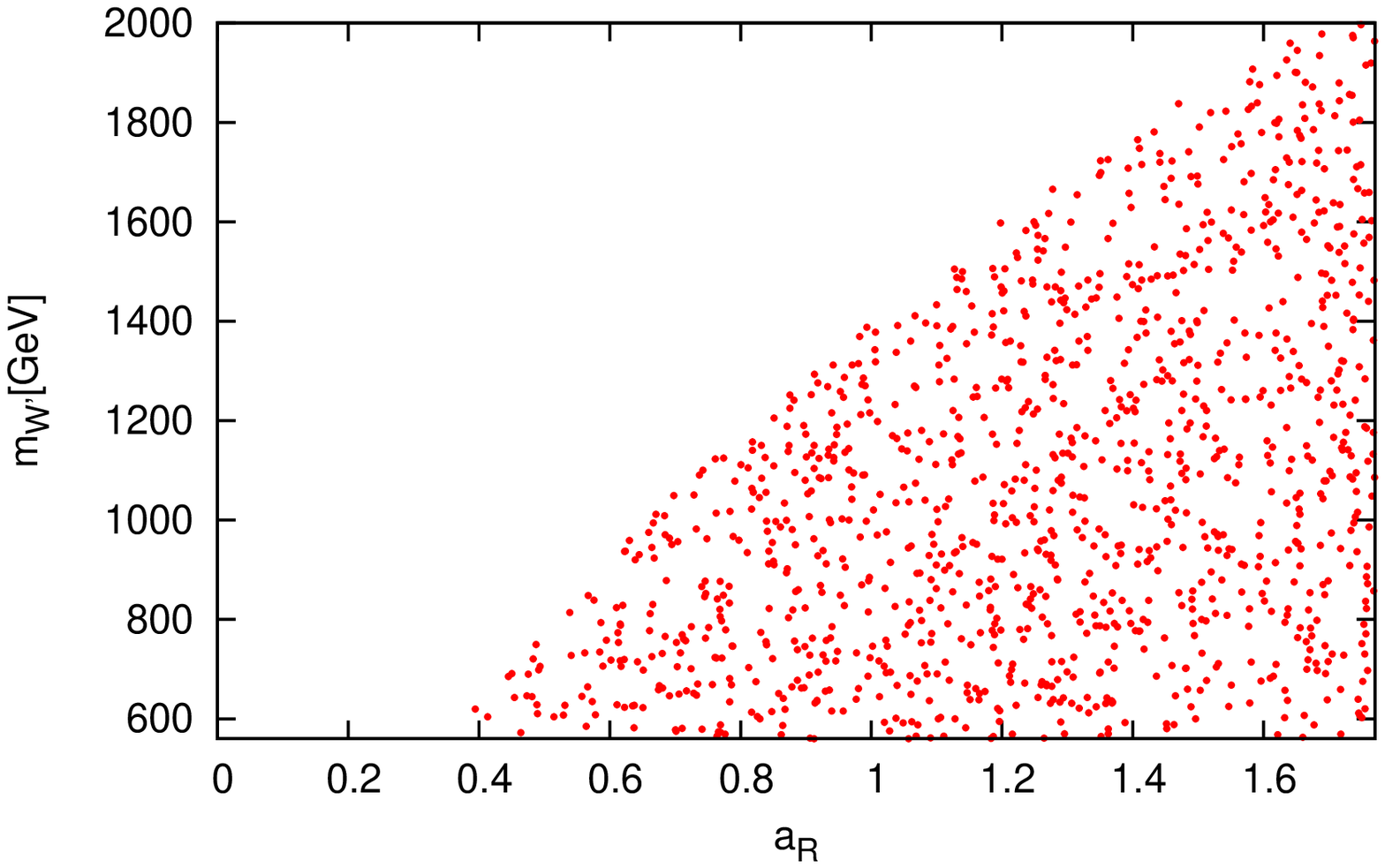,width=6.5cm}}
		\centerline{\vspace{-1.5cm}\hspace{0.5cm}(a)\hspace{6cm}(b)}
		\centerline{\vspace{-0.0cm}}
	\end{center}
	\caption{a) Scatter points depict regions in  a)$a_L$ ((b) $a_R$) and the $M_{W'}$ plane which are excluded by ATLAS measurement.}\label{fig:scatercross}
\end{figure}

In Fig.~\ref{fig:totalcross1}, 
we have shown the effect of \wprime exchange for $a_L=0$ and non-zero $a_R$ on single top quark production. 
As it is expected, with increasing $M_{W'}$, deviation from SM
cross section decreases. Figure \ref{fig:totalcross} (\ref{fig:totalcross1}) shows that the effect of presence of \wprime for $M_{W'}<1160\rm GeV$ ($M_{W'}<1390\rm GeV$) can be  larger than the ATLAS measurement. This means left-handed (right-handed) \wprime have been excluded for $M_{W'}<1160~\rm GeV$ ($M_{W'}<1390~\rm GeV$). Figure \ref{fig:scatercross} demonstrates this effect in the \wprime parameter space.
Scatter points depict regions in (a) $a_L$ ((b) $a_R$)  and the $M_{W'}$ plane which are excluded by ATLAS measurement. According to Fig.~\ref{fig:totalcross}-a (Fig.~\ref{fig:totalcross1}), 
the best exclusion limit on $M_{W'}$ for $a_L=1$ and $a_R=0$ ($a_R=1$ and $a_L=0$) is $1160~\rm GeV$ ($1390~\rm GeV$). As an alternative approach in the next section, we use a statistical method to constrain the \wprime phase space.

\section{Statistical Analysis}
In this study, we use the reported event yields by the experiments in search for s-channel single top production and find the upper limit on the contribution of the \wprime boson. In this part, the interference is neglected even for the left-handed \wprime boson. The data used in this study include the following categories:
\begin{description}
	\item[CMS $\mu$ 7 TeV] The search by the CMS experiment for the s-channel single top production in the events containing one isolated muon in data from the pp collisions at $\sqrt{s}$ = 7 TeV \cite{Khachatryan:2016ewo}.
	\item[CMS $\mu$ 8 TeV] The same search in data from the pp collisions at $\sqrt{s}$ = 8 TeV.
	\item[CMS electron 8 TeV] The same search when the isolated lepton is an electron.
	\item[ATLAS 8 TeV] The search by the ATLAS experiment for the s-channel single top production in the events containing one isolated electron or muon \cite{Aad:2015upn}. The ATLAS experiment has provided the numbers only for the sum of electron and muon in pp collisions at $\sqrt{s}$ = 8 TeV.
\end{description}
More data is available \cite{Abazov:2013qka,Aaltonen:2015xea,ATLAS:2011aia}, but it was tested and confirmed that adding them can not improve the results. It is also checked that the used categories of data are important and removing any of them can decrease the exclusion power. Table \ref{tab:Data} 
\begin{table}[!htb]
	\caption{Different categories of data used for the analysis. The reported uncertainties are the quadratic sum of the statistical and systematic uncertainties.  \label{tab:Data} }
	\begin{tabular}{|l|c|c|c|c|}
		\hline
		Center of mass energy &  7 TeV  & \multicolumn{3}{c|}{8 TeV}\\\hline
		Category              &  CMS $\mu$   & CMS $\mu$    & CMS electron & ATLAS\\\hline
		Single top s-channel  &  129 $\pm$ 5 & 452 $\pm$ 16 & 347 $\pm$ 12 & 540 $\pm$ 160 \\
		SM Backgrounds        &1920 $\pm$ 110&7060 $\pm$ 370&6240 $\pm$ 320& 14670 $\pm$ 180\\
		Observed data         &     1883     &     7023     & 6301         & 14677\\\hline
	\end{tabular}
\end{table}
summarizes the data used for this analysis. The quadratic sum of the statistical and systematic uncertainties are reported as the total uncertainty. The main part of the systematic uncertainty comes from the uncertainty on the jet energy scale.

For statistical calculations, the tools provided by the ROOT \cite{Brun:1997pa} data analysis framework are used. The efficiency of event reconstruction and selection for the events from \wprime contribution is assumed to be the same as the efficiency of the s-channel events. The SM cross section of the s-channel single top production in different data categories is taken from the corresponding analysis. The cross section of \wprime production with $tb$ final state can be found in Ref.\cite{Duffty:2012rf}. The cross sections are provided up to  next-to-leading order (NLO) of QCD precision. The relative uncertainty of the signal yield is assumed to be 30\% for all categories which is a conservative value. Trying other values like 20 or 40\% do not affect the final exclusions.

\begin{figure}[htbp]
	\centering 
	\includegraphics[width=.475\textwidth]{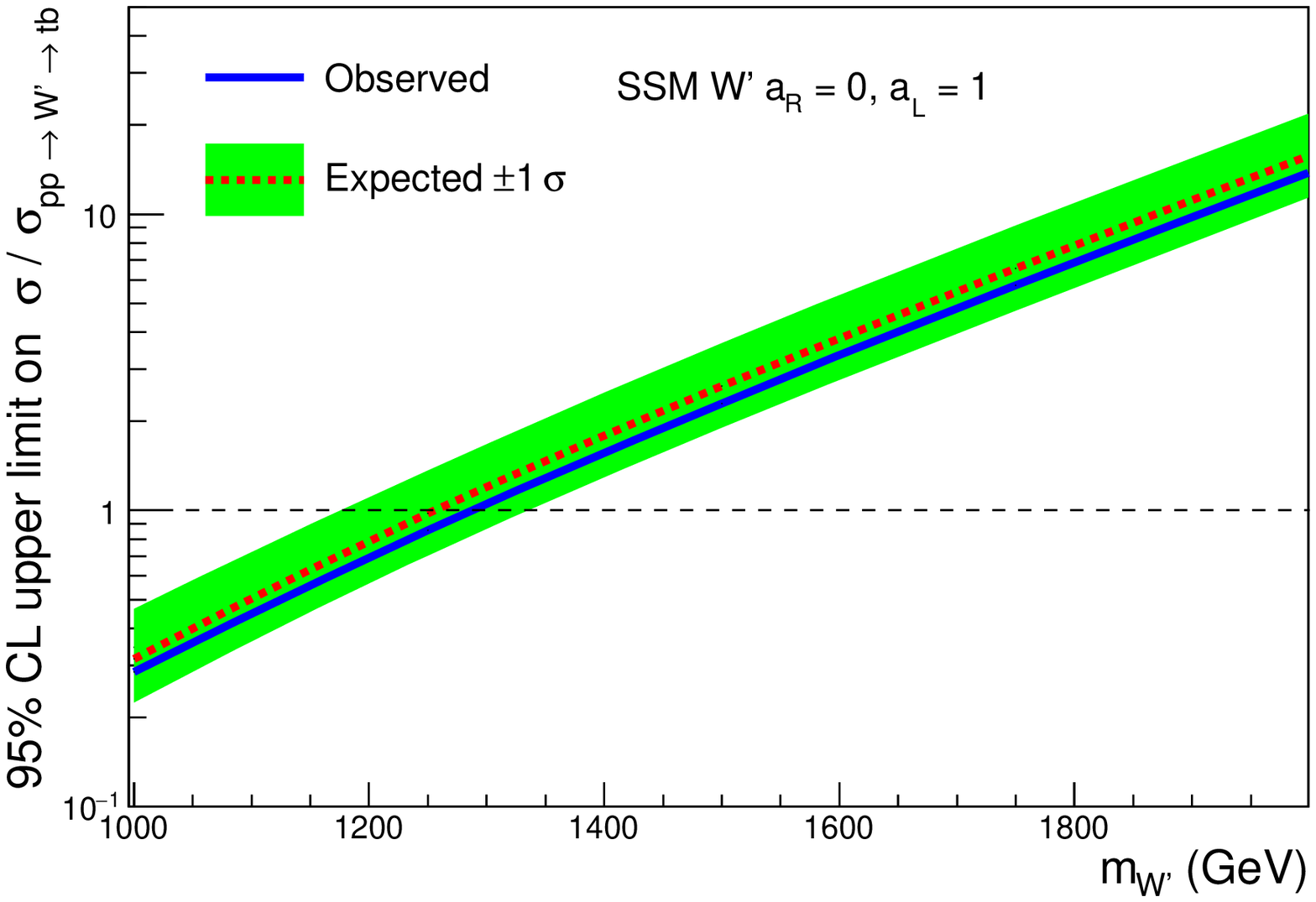}
	\hfill
	\includegraphics[width=.475\textwidth]{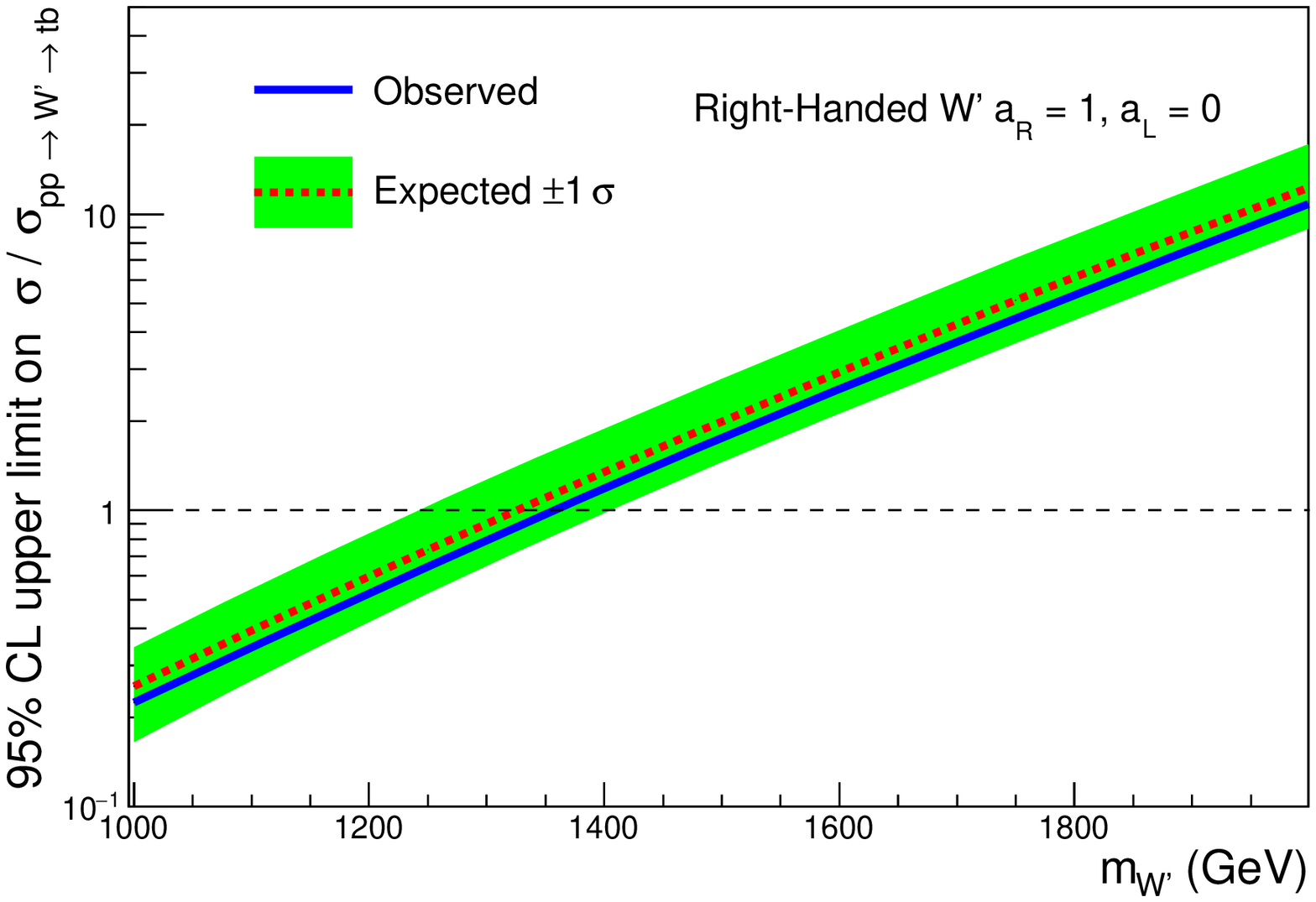}
	\caption{\label{fig:exclusion} The expected and observed limits on the cross section of \wprime decaying to tb.}
\end{figure}

Figure \ref{fig:exclusion} (left)
shows the 95\% CL upper limit on the contribution of the \wprime to the production cross section of the s-channel single top quark, when the \wprime is fully left-handed with the couplings similar to those of the SM W boson. It can be seen that \wprime boson with masses below 1290 GeV are excluded by the data. The expected value of the exclusion is 1255 GeV. The $\pm1$ $\sigma$ of the expected limits are shown as a green band. In Fig.~\ref{fig:exclusion} (right), the same information is presented for the fully right-handed \wprime boson, where the new boson does not couple to leptons. The expected (observed) limit on the mass of \wprime in this model is 1325 (1360) GeV. The results can be compared to the CMS limits based on 8 TeV data in the similar final state which is 2.15 TeV \cite{Khachatryan:2015edz}.

\begin{figure}[htbp]
	\centering 
	\includegraphics[width=.475\textwidth]{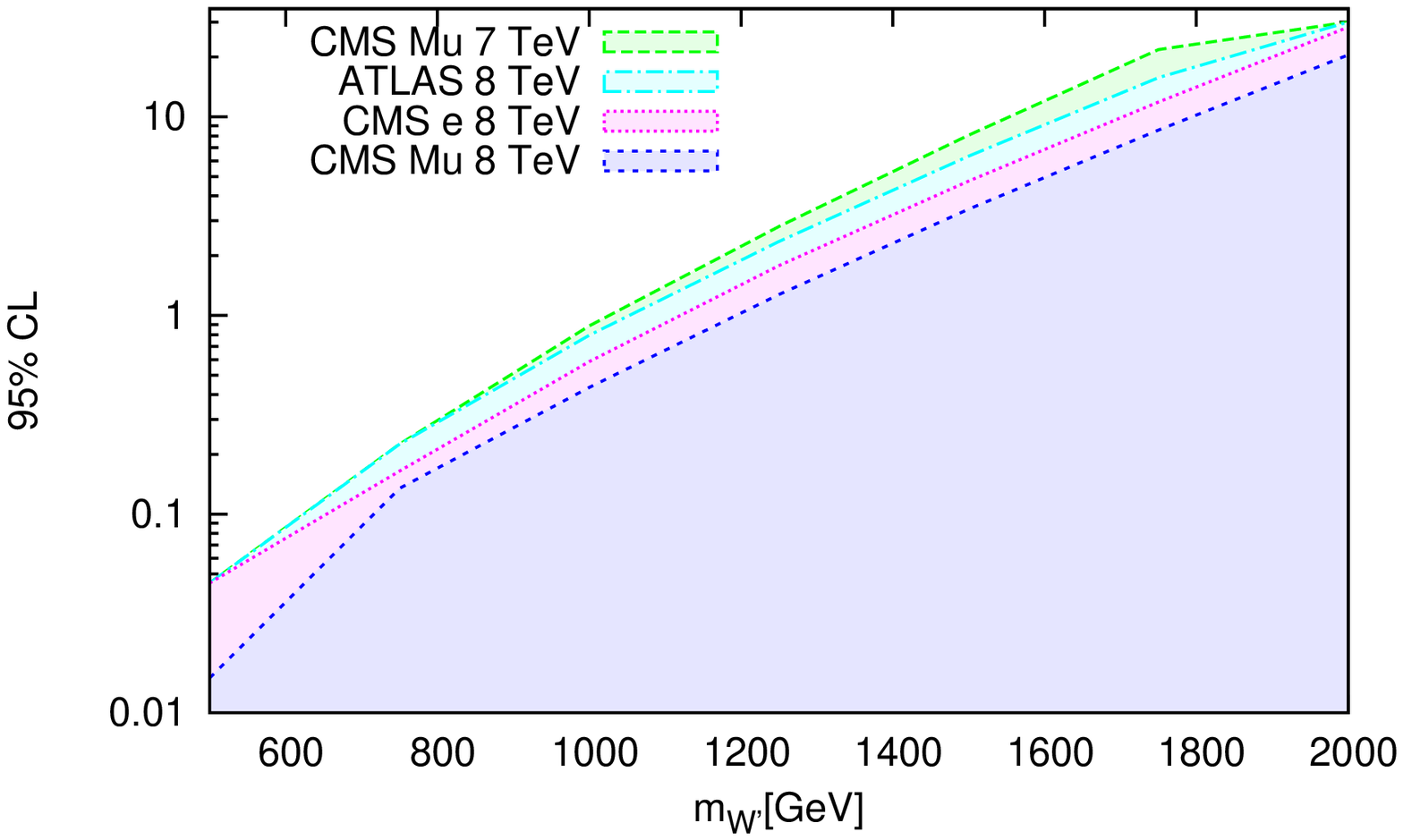}
	\hfill
	\includegraphics[width=.475\textwidth]{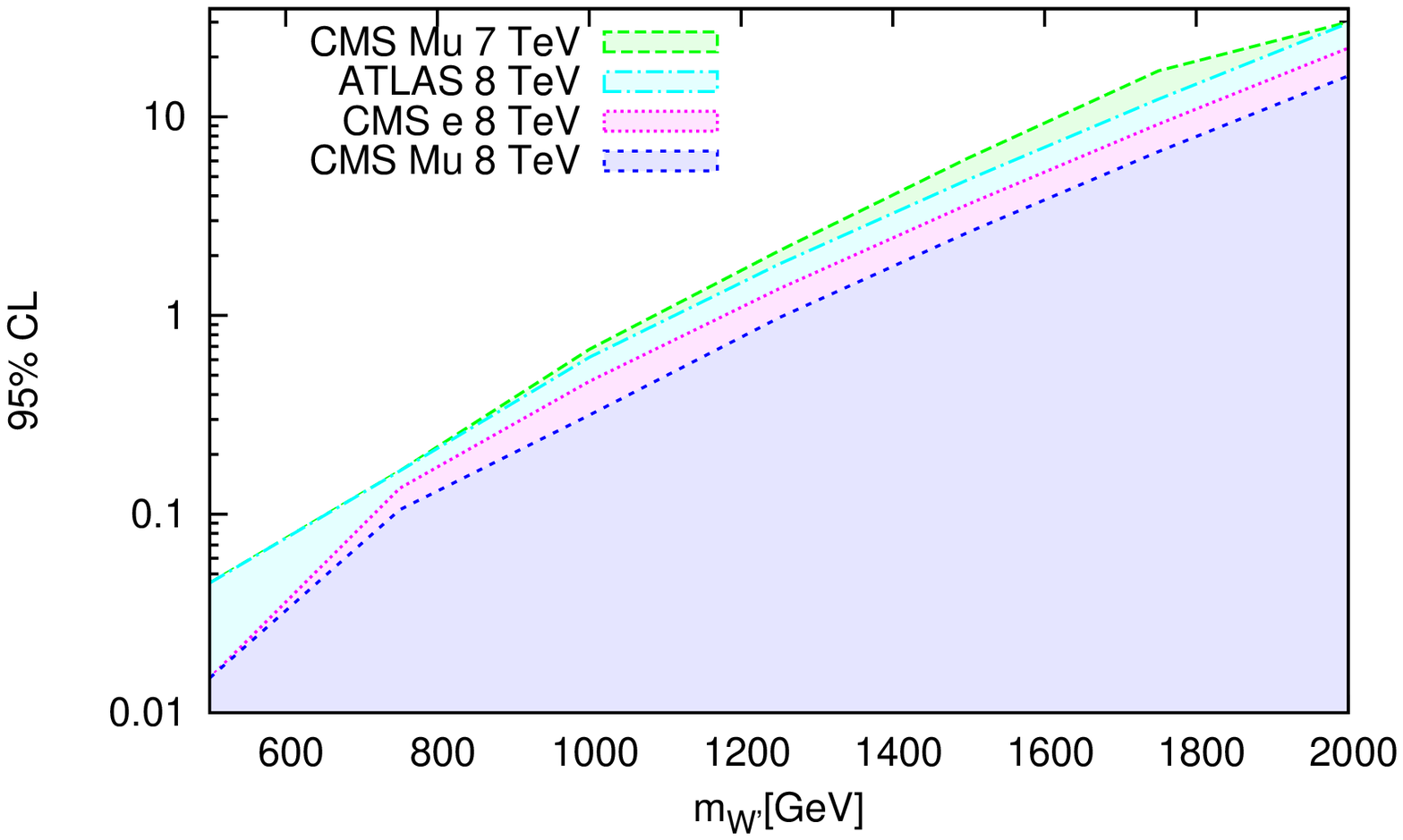}
	\caption{\label{fig:exclusionDetail} The observed limits for different categories of data. In the left(right), \wprime is pure left(right)-handed.}
\end{figure}

In Fig.~\ref{fig:exclusionDetail}, 
the observed 95\% CL upper limit is shown for different categories of data, separately. Although the ATLAS result is the most precise measurement for the s-channel single top production, but the main power for exclusions comes from "CMS $\mu$ 8 TeV". 
For the SSM \wprime, using this category alone can rule out the \wprime lighter than 1180 GeV. The power of the CMS results come from the detailed values for different channels, but the ATLAS experiment has reported only the sum of the yields of the electron and muon channels. In table \ref{tab.exclusionObsExp}
\begin{table}[!htb]
	\centering
	\caption{\label{tab.exclusionObsExp} The expected and observed limits driven by different categories of data.}
	\begin{tabular}{|c|cc|cc|}
		\hline
		& \multicolumn{2}{c|}{Left-handed} & \multicolumn{2}{c|}{Right-handed}\\
		dataset           & Expected & Observed & Expected & Observed \\\hline
		CMS $\mu$ 7 TeV       &   1000   &   1020   &   1050   &   1070   \\   
		CMS $\mu$ 8 TeV       &   1150   &   1180   &   1220   &   1250   \\   
		CMS electron 8 TeV    &   1090   &   1110   &   1160   &   1170   \\   
		ATLAS 8 TeV           &   1045   &   1045   &   1110   &   1120   \\\hline
	\end{tabular}
\end{table}
the expected and observed limits are shown for different categories of data, in pure left-handed or right-handed scenario. The data from pp collisions at $\sqrt{s}$ = 7 TeV give the worst limit and data from 8 TeV collisions have improved the limits significantly. One can hope that when the data from pp collisions at $\sqrt{s}$ = 13 TeV are available, the limits can be improved even more, keeping in mind that the volume of the new data is much more than the previous data also.

\begin{figure}[htbp]
	\centering 
	\includegraphics[width=.475\textwidth]{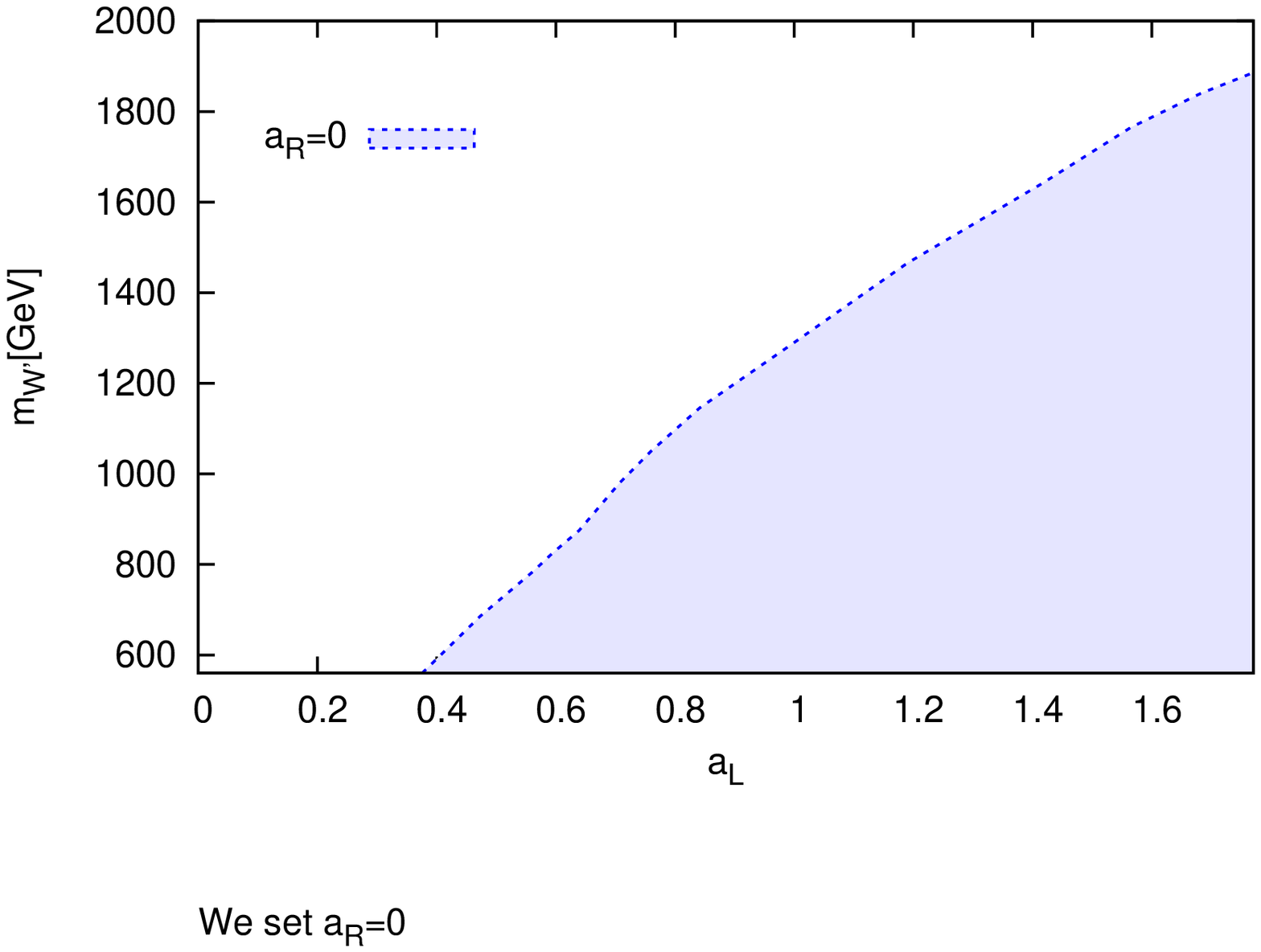}
	\hfill
	\includegraphics[width=.475\textwidth]{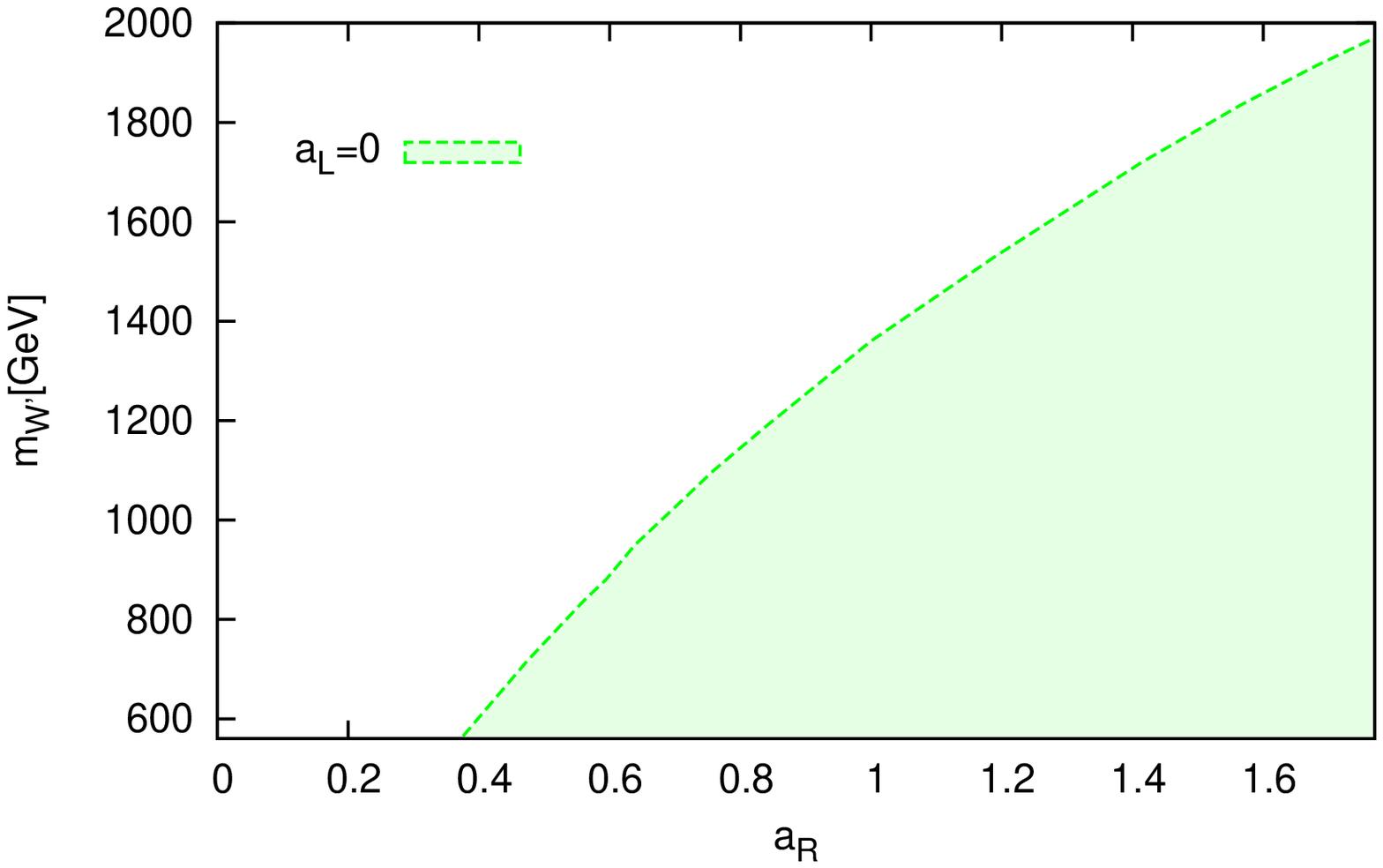}
	\caption{\label{fig:exclusionCouplingMass} Exclusion in the plane of \wprime mass and coupling. The shaded area is excluded by this analysis.}
\end{figure}

The same analysis can set limits on the coupling versus \wprime mass. In this part also, \wprime is either fully right-handed or fully left-handed and the mixture is not allowed.
Figure \ref{fig:exclusionCouplingMass} 
shows the region in the \wprime mass versus the coupling that can be excluded in 95\% CL. It can be seen that varying the $a_{L}$ ($a_{R}$) from 0.4 to 1.8 can exclude the \wprime mass between 600 and 1900 (2000) GeV. The main assumption here is the ($V-A$) nature of \wprime and couplings are free to change.
The figure can be compared to Fig.~\ref{fig:scatercross}, where the scattered points show the part of the phase space that produce the cross sections inconsistent with the  measured cross section of the s-channel single top production from the ATLAS experiment.

\section{Conclusion}
The heavy partner of $W$ boson, known as \wprime boson, can contribute to s-channel 
production of the single top quark. In this analysis, for the first time, the measured 
cross section of the s-channel single top quark is used to constrain the contribution 
of \wprime boson. In a phenomenological approach, the part of the parameter space which is excluded by
the measurement of the cross section  is found. 
In an alternative approach, the reported yields from the experiments used to measure the cross section
are statistically analyzed to set the upper limit on the yields of 
the \wprime boson events.  The latter analysis rules out the right-handed \wprime boson with masses 
below 1.36 TeV, while the former can push the limit up to 1.39 TeV. The limits for a left-handed \wprime 
boson are 1.16 TeV and 1.29 TeV from the phenomenological and statistical analysis, respectively. 
In both approaches, the excluded region in the plane of the coupling versus the \wprime boson mass is reported.

\section{Acknowledgment}
The authors would like to thank the school of particles and accelerators at IPM for their hospitality.

%

\end{document}